\documentclass[letterpaper,preprintnumbers,prd,twocolumn,nofootinbib,nobibnotes,showpacs]{revtex4}
\usepackage{epsfig}
\usepackage{graphicx}%
\usepackage{dcolumn}
\usepackage{amsmath}

\makeatletter
\def\btt#1{\texttt{\@backslashchar#1}}%
\DeclareRobustCommand\bblash{\btt{\@backslashchar}}%
\makeatother

\begin{document}

\title{Phantom Black Holes}
\author{Chang Jun Gao$^1$}\email{gaocj@mail.tsinghua.edu.cn}\author{Shuang Nan Zhang$^{1,2,3,4}$}
\email{zhangsn@mail.tsinghua.edu.cn}\affiliation{$^1$Department of
Physics and Center for Astrophysics, Tsinghua University, Beijing
100084, China(mailaddress)} \affiliation{$^2$Physics Department,
University of Alabama in Huntsville, AL 35899, USA}
\affiliation{$^3$Space Science Laboratory, NASA Marshall Space
Flight Center, SD50, Huntsville, AL 35812, USA}
\affiliation{$^4$Key Laboratory of Particle Astrophysics,
Institute of High Energy Physics, Chinese Academy of Sciences,
Beijing 100039, China}

\date{\today}

\begin{abstract}
The exact solutions of electrically charged phantom black holes
with the cosmological constant are constructed. They are labelled
 by the mass, the electrical charge, the cosmological constant and the coupling constant
 between the phantom and the Maxwell field. It is found that the phantom has important
 consequences on the properties of black holes. In particular, the extremal charged phantom black holes can never
 be achieved and so the third law of thermodynamics for black holes still
holds. The cosmological aspects of the phantom black hole and
phantom field are also briefly discussed.
\end{abstract}

\pacs{04.20.Ha, 04.50.+h, 04.70.Bw}
 \maketitle
\section{Introduction}
New analysis of SN Ia observations favor the parameter of the
equation of state for the dark energy with $w<-1$ at $1 \sigma$
level
 \cite{1}. Since the parameter of the equation of state of conventional
quintessence models with positive kinetic energy can not evolve to
the regime of $w<-1$, some authors
\cite{2,3,4,5,6,7,8,9,10,11,12,13,14,15,16,17} have investigated
the phantom field models that possess negative kinetic energy and
can achieve $w<-1$. As a candidate of dark energy, phantom field
contributes a repulsive force on the large structure of the
Universe and accelerates the expansion of the Universe. The action
of the phantom field is assumed to be
\begin{eqnarray}
S&=&\int{d^nx\sqrt{-g}}\left[R+\frac{4}{n-2}\partial_{\mu}\psi\partial^{\mu}{\psi}
-V\left(\psi\right)\right],
\end{eqnarray}
where $\psi$ is the phantom field and the scalar function
$V(\psi)$ is the phantom potential. Compared to the ordinary
scalar field, the action has only a sign difference before the
kinetic term. As far as we know, the explicit expression of the
phantom potential and the exact solution of black holes in the
phantom field (We call them phantom black holes) have not yet been
given. The goal of this paper is to find an explicit expression of
the phantom potential and also the exact solutions of the phantom
black holes.
\section{Higher dimensional and topological phantom black holes}
Let us start from an n-dimensional theory in which
 gravity is coupled to dilaton and Maxwell field with an action
\begin{eqnarray}
S&=&\int{d^nx\sqrt{-g}}\left[R-\frac{4}{n-2}\partial_{\mu}\phi\partial^{\mu}{\phi}
-V\left(\phi\right)\right.\nonumber
\\ &&\left.  -e^{-\frac{4\alpha\phi}{n-2}}F^2\right],
\end{eqnarray}
where $R$ is the scalar curvature, $F^2=F_{\mu\nu}F^{\mu\nu}$ is
the usual Maxwell contribution, $\alpha$ is an arbitrary constant
governing the strength of the coupling between the dilaton and the
Maxwell field, and $V\left(\phi\right)$ is a potential of dilaton
$\phi$ and corresponds to the cosmological constant which is given
by \cite{18}
\begin{eqnarray}
&&V\left(\phi\right)=\frac{\lambda}{3\left(n-3+\alpha^2\right)^2}
\nonumber\\&&\cdot\left[-\alpha^2\left(n-2\right)\left(n^2-n\alpha^2-6n+\alpha^2+9\right)
e^{-\frac{4\left(n-3\right)\left(\phi-\phi_0\right)}{\left(n-2\right)\alpha}}\right.\nonumber
\\ &&\left. +\left(n-2\right)
\left(n-3\right)^2\left(n-1-\alpha^2\right)e^{\frac{4\alpha\left(\phi-\phi_0\right)}{n-2}}\right.\nonumber
\\ &&\left. +4\alpha^2\left(n-3\right)
\left(n-2\right)^2e^{\frac{-2\left(\phi-\phi_0\right)\left(n-3-\alpha^2\right)}{\left(n-2\right)\alpha}}\right].
\end{eqnarray}
Here $\lambda$ is the cosmological constant and $\phi_0$ is the
asymptotic value of dilaton which can be absorbed by $\phi$. One
can verify that the potential reduces to the Einstein cosmological
constant when $\alpha=0$ or $\phi=0$. We should point out if and
only if by using this potential, can we obtain the asymptotically
de Sitter dilaton black hole solutions. Thus it is the counterpart of Einstein cosmological constant. \\
\hspace*{3.5mm}Compared to the action of the ordinary scalar
fields, the phantom field has the negative kinetic term. In order
to obtain a real action of the Einstein-Maxwell field in the
presence of the phantom, we can make a mathematical trick, the
so-called Wick rotation, in the action while without thinking the
physical meaning as follows
\begin{equation}
\phi\rightarrow i\psi, \ \ \ \ \alpha\rightarrow i\beta,
\end{equation}
where $i$ is the imaginary unit. Then we get the action
\begin{eqnarray}
S&=&\int{d^nx\sqrt{-g}}\left[R+\frac{4}{n-2}\partial_{\mu}\psi\partial^{\mu}\psi-V\left(\psi\right)
\right.\nonumber
\\ &&\left. -e^{\frac{4\beta\psi}{n-2}}F^2\right],
\end{eqnarray}
and the potential for the phantom field
\begin{eqnarray}
&&V\left(\psi\right)=\frac{\lambda}{3\left(n-3-\beta^2\right)^2}
\nonumber\\&&\cdot\left[\beta^2\left(n-2\right)\left(n^2+n\beta^2-6n-\beta^2+9\right)
e^{-\frac{4\left(n-3\right)\psi}{\left(n-2\right)\beta}}\right.\nonumber
\\ &&\left. +\left(n-2\right)
\left(n-3\right)^2\left(n-1+\beta^2\right)e^{-\frac{4\beta\psi}{n-2}}\right.\nonumber
\\ &&\left. -4\beta^2\left(n-3\right)
\left(n-2\right)^2e^{\frac{-2\psi\left(n-3+\beta^2\right)}{\left(n-2\right)\beta}}\right].
\end{eqnarray}
We note that $\psi_0$ has been absorbed by $\psi$. One can also
verify that, when $\psi=0$ or $\beta=0$ the action reduces to the
Einstein-Maxwell action and when $F_{\mu\nu}=0$ the action reduces
to the Einstein-phantom action. It is apparent that changing the
sign of $\beta$ is equivalent to changing the sign of $\psi$.
Thus it is sufficient to consider only $\beta > 0$. \\
\hspace*{3.5mm}Using the above method of variables substitutions
of Eq.(4) we can immediately write down the metrics of the phantom
black holes with cosmological constant in contrast to the dilaton
version \cite{19},
\begin{eqnarray}
d{s}^2&=&-\left\{\left[k-\left(\frac{r_{+}}{{r}}\right)^{n-3}\right]
\left[1-\left(\frac{r_{-}}{{r}}\right)^{n-3}\right]^{1-\gamma\left(n-3\right)}
\right.\nonumber
\\ &&\left. -\frac{1}{3}\lambda
r^2\left[1-\left(\frac{r_{-}}{{r}}\right)^{n-3} \right]^{\gamma}
\right\}d{t}^2\nonumber\\&&+\left\{\left[k-\left(\frac{r_{+}}{{r}}\right)^{n-3}\right]
\left[1-\left(\frac{r_{-}}{{r}}\right)^{n-3}\right]^{1-\gamma\left(n-3\right)}
\right.\nonumber
\\ &&\left. -\frac{1}{3}\lambda
r^2\left[1-\left(\frac{r_{-}}{{r}}\right)^{n-3} \right]^{\gamma}
\right\}^{-1}\nonumber\\
&&\cdot\left[1-\left(\frac{r_{-}}{{r}}\right)^{n-3}
\right]^{-\gamma\left(n-4\right)}d{r}^2\nonumber\\&&+
r^2\left[1-\left(\frac{r_{-}}{{r}}\right)^{n-3}\right]^{\gamma}d\Omega_{k,n-2}^2,
\end{eqnarray}
where $r_+$ and $r_{-}$ are the two horizons of the black hole,
and $\gamma$, physical mass $M$ and electrical charge $Q$ are
given by
\begin{eqnarray}
\gamma &=&\frac{-2\beta^2}{\left(n-3\right)\left(n-3-\beta^2\right)},\nonumber\\
Q^2&=&\frac{\left(n-2\right)\left(n-3\right)^2}{2\left(n-3-\beta^2\right)}
r_+^{n-3}r_-^{n-3},\nonumber\\
M&=&\frac{r_+}{2}\left(n-3\right)\left[1-\left(\frac{r_{-}}{r_+}\right)^{n-3}\right]
^{\frac{\left(n-3\right)^2-\beta^2}{\left(n-3\right)\left(n-3-\beta^2\right)}}
\nonumber\\&&+\frac{\left(n-2\right)\left(n-3\right)}{2\left(n-3-\beta^2\right)}r_{-}^{n-3}.
\end{eqnarray}
$k=0, \pm 1$ denotes the three kinds of topologies of black holes.
For $k=1$, the spacetime has the topology of $R^2\bigotimes
S^{n-2}$, i.e., the horizons of the black hole have the topology
of a $(n-2)$ dimensional sphere. For $k=0$, the spacetime has the
topology of $R^2\bigotimes S^1\bigotimes S^{n-3}$ by identified
$\varphi=0$ with $\varphi=2\pi$ and $\theta=0$ with $\theta=\pi$,
i.e., the horizons of the black hole have the topology of a
$(n-2)$ dimensional torus. For $k=-1$, the spacetime has the
topology of $R^2\bigotimes R^1\bigotimes S^{n-3}$ also  by
identified $\varphi=0$ with $\varphi=2\pi$, i.e., the horizons of
the black hole have the topology of a $(n-2)$ dimensional
hyperboloid.
\section{four dimensional and spherical black holes}
As an example, we focus on the four dimensional and spherical
solution. The cosmological constant is also omitted. Then the
metric is given by
\begin{eqnarray}
d{s}^2&=&-\left(1-\frac{r_{+}}{{r}}\right)\left(1-\frac{r_{-}}{{r}}\right)^{\frac{1+\beta^2}{1-\beta^2}}
d{t}^2\nonumber\\&&+\left(1-\frac{r_{+}}{{r}}\right)^{-1}\left(1-\frac{r_{-}}{{r}}\right)^{-\frac{1+\beta^2}{1-\beta^2}}d{r}^2
\nonumber\\
&&+r^2\left(1-\frac{r_{-}}{r}\right)^{\frac{-2\beta^2}{1-\beta^2}}d\Omega_{2}^2,
\end{eqnarray}
and the phantom field, physical mass and electrical charge of the
black hole are given by
\begin{eqnarray}
e^{-2\beta\psi}&=&\left(1-\frac{r_{-}}{r}\right)^{\frac{-2\beta^2}{1-\beta^2}},\
\ \ \ Q^2=\frac{r_+r_{-}}{1-\beta^2},\nonumber\\
M&=&\frac{r_+}{2}+\frac{1+\beta^2}{1-\beta^2}\cdot\frac{r_{-}}{2}.
\end{eqnarray}
\hspace*{3.5mm}When $\beta=0$, the solution reduces to the
Reissner-Nordstr$\ddot{\textrm{o}}$m solution. However, for
$\beta\neq0$ the solution is qualitatively different. For all
$\beta$, $r=r_{+}$ is an event horizon. The surface $r=r_{-}$ is a
curvature singularity except for the case $\beta=0$ when it is a
nonsingular inner horizon. Thus they describe black holes only
when $r_{-}<r_{+}$. Then the two horizons $r_{+}$ and $r_{-}$
locate, respectively, at
\begin{eqnarray}
r_{+}&=& M+\sqrt{M^2-\left(1+\beta^2\right)Q^2},\nonumber\\
r_{-}&=&\frac{1-\beta^2}{1+\beta^2}\left[M-\sqrt{M^2-\left(1+\beta^2\right)Q^2}\right].
\end{eqnarray}
For $\beta\gg1$, Eqs.(11) tell us that a small amount of
electrical charge would produce a large change in the geometry
close to the horizon. For $\beta\neq 0$, the extremal black hole
$r_+=r_{-}$ can never be achieved. The surface gravity is
\begin{eqnarray}
\kappa=\frac{1}{2r_{+}}\left(1-\frac{r_{-}}{r_{+}}\right)^{\frac{1+\beta^2}{1-\beta^2}}.
\end{eqnarray}
Thus the surface gravity will never approach zero except for
$\beta=0$. For all $\beta$, it does not diverge. Since the
temperature is proportional to $\kappa$, the third law of
thermodynamics for black holes still holds. This is very much
different from the dilaton case where for $\beta<1$ the surface
gravity goes to zero in the extremal limit,
 for $\beta=1$ it approaches a constant and for $\beta>1$ it diverges.\\
\hspace*{3.5mm}We note that the phantom black holes have several
other significant differences from the dilaton ones \cite{20}. In
the first place the transition between black holes and naked
singularities occurs at $Q={M}/{\sqrt{1+\beta^2}}$ in Eqs.(11)
rather than $Q={M}/{\sqrt{1-\beta^2}}$ as in the dilaton case. In
other words, to achieve a naked singularity, we need a smaller
charge compared to the dilaton case. This can be understood as
follows. For the electrically charged dilaton black holes, the
extremal value corresponds to the case where the repulsive force
of the electric charge can exactly destroy the event horizon (or
the repulsive force of electric charge exactly balances the
attractive forces of mass and dilaton). However, for the phantom
black holes, the extremal value corresponds to the case where the
repulsive forces of electrical charge and phantom charge can
exactly destroy the event horizon. In other words, dilaton field
contributes an extra attractive force and phantom field
contributes an extra repulsive force between black holes. So for a
given $M$, one needs a smaller $Q$ to destroy the event horizon.
We will return to this point in section $\textrm{\textsc{VI}}$.
Secondly, the curvature singularity is present for all $\beta$ for
dilaton black hole. In contrast, it is present only for
$0\leq\beta<1$ in the phantom case.
\section{a specific case of the coupling constant}
In the next for brevity but without the loss of generality, we
consider the case of $\beta=1$. It is found that it is of
particular interest. For $\beta=1$, the metric becomes
\begin{eqnarray}
d{s}^2&=&-\left(1-\frac{r_{+}}{{r}}\right)e^{\frac{-r_{-}}{{r}}}
d{t}^2+\left(1-\frac{r_{+}}{{r}}\right)^{-1}e^{\frac{r_{-}}{{r}}}d{r}^2\nonumber\\&&
+r^2e^{\frac{r_{-}}{{r}}}d\Omega_{2}^2,\nonumber\\
 r_+&=&M+\sqrt{M-2Q^2},\nonumber\\ r_-&=&M-\sqrt{M-2Q^2}.
\end{eqnarray}
The phantom charge is given by
\begin{eqnarray}
P=\frac{1}{4\pi}\int
d^2\Sigma^{\mu}\nabla_{\mu}\psi=\frac{1}{2}\left(\sqrt{M^2-2Q^2}-M\right).
\end{eqnarray}
It is not a new parameter and is determined by its mass and
charge. It is easy to find that $P$ is in the range of $[-M/2,
0]$. The metric of Eqs.(13) and the phantom field can be rewritten
as
\begin{eqnarray}
d{s}^2&=&-\left(1-\frac{2M+2P}{{r}}\right)e^{\frac{2P}{{r}}}
d{t}^2\nonumber\\&&+\left(1-\frac{2M+2P}{{r}}\right)^{-1}e^{-\frac{2P}{{r}}}d{r}^2
+r^2e^{-\frac{2P}{{r}}}d\Omega_{2}^2,\nonumber\\
\psi&=&\frac{P}{r}.
\end{eqnarray}
When $P=0$, it reduces to the Schwarzschild solution. We recall
that the Newtonian gravitational field with mass $M$ is
$\psi_N=-M/r$. Eqs.(15) tell us the phantom field with charge $P$
is $\psi_P=P/r$ (P is negative). However, we can not say the
phantom charge contributes a long-range, attractive force to the
physical mass. This can be understood from the following example.
We know that the Coulomb field with electrical charge $Q$ is
$\psi_Q=Q/r$. However, we can not say that the electrical charge
contributes a long-range, repulsive force to the physical mass.\\
\hspace*{3.5mm}Similar to the Schwarzschild case, $r=2M+2P$ is the
regular event horizon and $r=0$ is the curvature singularity. The
corresponding Hawking temperature is
\begin{eqnarray}
T=\frac{e^{-\frac{P}{M+P}}}{8\pi\left(M+P\right)}.
\end{eqnarray}
This reveals that the corresponding Hawking temperature increases
with the presence of the phantom charge. For the maximum value of
electrical charge, ie. $Q=M/\sqrt{2}$ (That is the transition
between black
hole and singularity), we have the non-vanishing temperature $T=e/(4\pi M)$.\\
\section{physical realization of the phantom black hole}
\hspace*{3.5mm}In this section, we will focus on the physical
realization of the phantom black hole in the ordinary
gravitational collapse process. To this end, let us look for the
interior solution of an electrically charged static fluid ball
which is immersed in the phantom scalar field. Namely, the content
of the ball includes fluid, Maxwell field and the phantom scalar
field. We require that the solution should smoothly matches the
phantom black hole solution, Eq.(15). The related physical
quantities should also be reasonable. The field equations which
describe the fluid ball can be written as
\begin{eqnarray}
0&=&\nabla^2\psi+\frac{1}{2}e^{2\psi}F^2,\nonumber\\
0&=&F_{[\mu\nu;\alpha]},\nonumber\\
4\pi J^{\nu}&=&\nabla_{\mu}\left(e^{2\psi}F^{\mu\nu}\right)
,\nonumber\\
G_0^0&=&8\pi\rho-\nabla_\alpha\psi\nabla^\alpha\psi-\frac{1}{2}e^{2\psi}F^2,\nonumber\\
G_1^1&=&-8\pi
p_1+\nabla_\alpha\psi\nabla^\alpha\psi-\frac{1}{2}e^{2\psi}F^2,\nonumber\\
G_2^2&=&-8\pi
p_2-\nabla_\alpha\psi\nabla^\alpha\psi+\frac{1}{2}e^{2\psi}F^2.
\end{eqnarray}
The first equation is for the phantom field $\psi$. The second and
the third ones are for Maxwell field $F_{\mu\nu}$. $J^{\nu}$ is
the flux density of the electrical charges. Since we are looking
for a static solution for the ball, both $F^{\mu\nu}$ and
$J^{\nu}$ have only one non-vanishing component. They are $F^{10}$
and $J^0$ which represent the electric-field intensity and the
electric-charge density. This choice automatically satisfies the
second equation. The last three are the Einstein equations.
$G_{\mu}^{\nu}, \rho, p_1, p_2$ denote, respectively, the Einstein
tensor, the matter density, the radial pressure and the tangent
pressure. We set the metric has the form
\begin{eqnarray}
ds^2&=&-e^{\gamma\left(r\right)}dt^2+e^{\lambda\left(r\right)}dr^2+f\left(r\right)^2d\Omega_2^2.
\end{eqnarray}
\hspace*{3.5mm}Now let's solve the above equations. We note that
we have nine variables to be determined, $\gamma, \lambda, f,
\psi, F^{10}, J^0, \rho, p_1, p_2$. However, since the second
equation in Eqs.(17) is automatically satisfied, we have only five
constraint equations. So we have four freedom to determine these
variables. Thus we may properly construct four functions in
advance, $\gamma, \lambda, f, \psi$. Then we will have five
unresolved functions and five constraint equations. So the problem
becomes complete. Reminded by the well-known Schwarzschild
interior solution
\begin{eqnarray}
ds^2&=&-\left(A-B\sqrt{1-2Mr^2/r_0^3}\right)^2dt^2\nonumber\\&&+
\left(1-2Mr^2/r_0^3\right)^{-1}dr^2+r^2d\Omega_2^2,
\end{eqnarray}
and the phantom black hole solution Eq.(15), we assume the charged
fluid ball have the form of
\begin{eqnarray}
ds^2&=&-\left(A-Be^{Pr^2/r_0^3}\sqrt{1-2\left(M+P\right)r^2/r_0^3}\right)^2dt^2\nonumber\\&&+
\left[1-2\left(M+P\right)r^2/r_0^3\right]^{-1}e^{-2Pr^2/r_0^3}dr^2
\nonumber\\&&+r^2e^{-2P/r_0^3r^2}d\Omega_2^2,
\end{eqnarray}
and assume the phantom scalar field is given by
\begin{eqnarray}
\psi&=&Pr^2/r_0^3,
\end{eqnarray}
where $A,B,r_0$ are three constants. Then the four variables
$\gamma, \lambda, f, \psi$ are defined. The next work is to solve
for $F^{10}, J^0, \rho, p_1, p_2$ by using Eqs.(17). It is found
that it is very easy. Due to the lengthy of the expressions of
$F^{10}, J^0, \rho, p_1, p_2$, they are not given here. We would
like to point out that Eq.(20) and
(21) are regular in the fluid ball and thus physical. \\
\hspace*{3.5mm}Now let's consider the matching conditions. In the
surface of the ball, $r=r_0$, the metric should smoothly matches
the phantom black hole one; the radial pressure becomes zero; the
phantom field is $\psi=P/r_0$ and the Maxwell field is
$F^2|_{r=r_0}=-2Q^2/r_0^4=4P^2/r_0^4+4PM/r_0^4$. These conditions
constitute the following equations
\begin{eqnarray}
-g_{00}|_{r=r0}&=&\left(1-\frac{2M+2P}{r_0}\right)e^{2P/r_0},\nonumber\\
g_{11}|_{r=r0}&=&\left(1-\frac{2M+2P}{r_0}\right)^{-1}e^{-2P/r_0},\nonumber\\
g_{22}|_{r=r_0}&=&r_0^2e^{-2P/r_0},\nonumber\\
p_1|_{r=r_0}&=&0,\nonumber\\
\psi|_{r=r_0}&=&P/r_0,\nonumber\\
F^2|_{r=r_0}&=&4P^2/r_0^4+4PM/r_0^4.
\end{eqnarray}
It is found that there are only three independent equations in
Eqs.(22). Thus we obtain
\begin{eqnarray}
A&=&\frac{3r_0 e^{P/r_0}}{\left(r_0-P\right)}\sqrt{\frac{P}{5P-2r_0}},\nonumber\\
B&=&\frac{r_0+2P}{2\left(r_0-P\right)},\nonumber\\
M&=&\frac{r_0^2-4Pr_0+5P^2}{2r_0-5P}.
\end{eqnarray}
We see that the fluid ball is described by only two parameters,
i.e. $M, P$ or $P, r_0$ or $M,r_0$. Namely, given the mass $M$ and
the radius $r_0$ of the ball, then the phantom charge $P$ is
constrained. This is required by the matching conditions. Using
above expressions, we have checked that the quantities $F^{10},
J^0, \rho, p_1, p_2$ are all physically reasonable. We find that
both the radial pressure and the tangent pressure increase when
approaching to the center of the ball. At $r=0$, they are given by
\begin{eqnarray}
p_1&=&\frac{\left(3B-A\right)M}{4\pi r_0^3\left(A-B\right)},\nonumber\\
p_2&=&\frac{\left(3B-A\right)M-6P\left(A-B\right)}{4\pi
r_0^3\left(A-B\right)}.
\end{eqnarray}
So they will become infinite when $A=B$ from which we obtain the
minimum radius $r_{min}$ for the ball to be stable.  This minimum
radius is bigger than the event horizon of the phantom black hole.
Thus we see that if we compress the mass $M$ and the phantom
charge $P$ within the radius $r_{min}$, there will need infinite
pressures to against the gravity. In other words, the ball will
collapse and form a phantom black hole. It is just like the
formation of Reissner-Nordstr$\ddot{\textrm{o}}$m black hole.
\section{phantom black hole in the FRW universe}
\hspace*{3.5mm}Because of the potential use of phantom black holes
in the evolution of the Universe, it is interesting to investigate
the phantom black holes in the background of
Friedmann-Robertson-Walker Universe. It is found that the metric
of a phantom black hole in de Sitter universe can be written as
\begin{eqnarray}
ds^2&=&-\frac{\left(1-\frac{M+P}{ax}\right)^2}{\left(1+\frac{M+P}{ax}\right)^2}
e^{\frac{4P}{ax}\left(1+\frac{M+P}{ax}\right)^{-2}}du^2\nonumber\\&&
+a^2{\left(1+\frac{M+P}{ax}\right)^4}
e^{\frac{-4P}{ax}\left(1+\frac{M+P}{ax}\right)^{-2}}\nonumber\\&&\cdot
\left(dx^2+x^2d\Omega_2^2\right),
\end{eqnarray}
where $a=e^{Hu}$. $H$ is the Hubble constant. If $M$ and $P$ are
put equal to zero, the metric reduces to the metric of the de
Sitter universe. If only $P$ is put equal to zero and $a$ is
assumed to be an arbitrary function of $u$, the metric reduces to
the Schwarzschild black hole in the flat FRW Universe \cite{21}.
When $H=0$, we have checked that it becomes the solution of an
isolated phantom black hole, namely, Eq.(15). In the following, we
will show Eq.(25) turns out to be
\begin{eqnarray}
d{s}^2&=&-\left[\left(1-\frac{2M+2P}{{r}}\right)e^{\frac{2P}{{r}}}-H^2r^2e^{-\frac{2P}{{r}}}\right]
d{t}^2\nonumber\\&&+\left[\left(1-\frac{2M+2P}{{r}}\right)e^{\frac{2P}{{r}}}-H^2r^2e^{-\frac{2P}{{r}}}\right]^{-1}d{r}^2
\nonumber\\&&+r^2e^{-\frac{2P}{{r}}}d\Omega_{2}^2,\nonumber\\
\end{eqnarray}
which can also be obtained from Eq.(7) by setting $n=4$ and
$\alpha=1$. To do this, set
\begin{eqnarray}
x&=&\frac{1}{a}\left(r-M-P+\sqrt{r^2-2Mr-2Pr}\right),
\end{eqnarray}
Eq.(25) reduces to
\begin{eqnarray}
d{s}^2&=&-\left[\left(1-\frac{2M+2P}{{r}}\right)e^{\frac{2P}{{r}}}-4H^2r^2e^{-\frac{2P}{{r}}}\right]
d{u}^2\nonumber\\&&-\frac{8Hr^2e^{-2P/r}}{\sqrt{r^2-2Mr-2Pr}}dudr
+\frac{4r^2e^{-2P/r}}{r^2-2Mr-2Pr}dr^2\nonumber\\&&+
4r^2e^{-\frac{2P}{{r}}}d\Omega_{2}^2.
\end{eqnarray}
To eliminate the cross term $dudr$, set once more
\begin{eqnarray}
du&=&dt-\left[\left(1-\frac{2M+2P}{{r}}\right)e^{\frac{2P}{{r}}}-4H^2r^2e^{-\frac{2P}{{r}}}\right]^{-1}
\nonumber\\&&\cdot\frac{4Hr^2e^{-2P/r}}{\sqrt{r^2-2Mr-2Pr}}dr.
\end{eqnarray}
So Eq.(28) can be written as
\begin{eqnarray}
d{s}^2&=&-\left[\left(1-\frac{2M+2P}{{r}}\right)e^{\frac{2P}{{r}}}-4H^2r^2e^{-\frac{2P}{{r}}}\right]
d{t}^2\nonumber\\&&+4\left[\left(1-\frac{2M+2P}{{r}}\right)e^{\frac{2P}{{r}}}-4H^2r^2e^{-\frac{2P}{{r}}}\right]^{-1}d{r}^2
\nonumber\\&&+4r^2e^{-\frac{2P}{{r}}}d\Omega_{2}^2.
\end{eqnarray}
Rescale the time coordinate $t$ and the Hubble constant $H$, then
the two metrics, Eq.(30) and Eq.(26), are identical. Thus we have
completed the verification of the metric in Eq.(25).\\
\hspace*{3.5mm}We note that the de Sitter universe is a special
case of FRW metric. For arbitrary function $a=a(u)$, Eq.(25) would
represent the phantom black hole in the background of FRW
universe. The detailed expressions of energy density and pressure
calculated from Eq.(25) are lengthy and tedious, so they are not
given here. In contrast, the energy flux density is relatively
simple
\begin{eqnarray}
J_P&=&\frac{a^3x^5P^2\left(ax-M-P\right)^2}{\pi\left(ax+M+P\right)^{10}}
e^{\frac{4Pax}{\left(ax+M+P\right)^2}}\frac{da}{du}.
\end{eqnarray}
It is apparent $J_P$ is closely related to the phantom charge $P$
and the evolution of $a$. Provided that $P$ is not zero, there
will be flow of matter as a whole either towards or away from the
black hole dependent on the evolution of $a$. For radiation field
$a\propto u^{{1/2}}$ or cold matter $a\propto u^{2/3}$, the flow
is always away from the black hole. The reason for this is that
the phantom charge contributes a repulsive force to physical mass.
On the contrary, we will see in the following the flow is always
towards the black hole for dilaton black hole. On the other hand,
when $ax\gg M+P$, we have
\begin{eqnarray}
J_P&\simeq&\frac{P^2}{\pi a^5x^3}\frac{da}{du}.
\end{eqnarray}
 Thus the flow of matter decreases to zero very
quickly in space. We remember that Eq.(25) also describes a
charged phantom object in the FRW universe. Now since the flow is
outwards all the time in the expanding Universe, we conclude that
at some times, the object would be evaporated completely. \\
\hspace*{3.5mm}In this regard, Babichev et al [22] have studied
the aspect of black holes in phantom fields, namely, the accretion
of phantom fluid onto a black hole. They have found a very
interesting feature, that the mass (and consequently the entropy)
of a black hole decreases in such a process which is similar to
our result here.\\
\hspace*{3.5mm}We would like to point out for the
Reissner-Nordstr$\ddot{o}$m black hole in FRW universe
\begin{eqnarray}
ds^2&=&
-\frac{\left[1-\frac{M^2}{a^2x^2}+\frac{Q^2}{a^2x^2}\right]^2}{\left[\left(1+\frac{M}{ax}
\right)^2 -\frac{Q^2}{a^2x^2}\right]^2}du^2\nonumber\\&&
+{a^2}\left[\left(1+\frac{M}{ax}\right)^2
-\frac{Q^2}{a^2x^2}\right]^2\nonumber\\&&\cdot\left(dx^2+x^2d\Omega_2^2\right),
\end{eqnarray}
there is no energy flux density. Thus even in the cosmological
aspect, the phantom charged black holes are very different from
the ordinary charged ones. As for the dilaton charged black hole
in the FRW universe
\begin{eqnarray}
ds^2&=&-\frac{\left(1-\frac{M}{ax}+\frac{D}{ax}\right)^2}{\left(1+\frac{M}{ax}+\frac{D}{ax}\right)^2
-\frac{4MD}{a^2x^2}}du^2\nonumber\\&&
+\left[\left(1+\frac{M}{ax}+\frac{D}{ax}\right)^2
-\frac{4MD}{a^2x^2}\right]\nonumber\\&& \cdot
a^2{\left(1+\frac{M}{ax}-\frac{D}{ax}\right)^2}\left(dx^2+x^2d\Omega_{2}^2\right),
\end{eqnarray}
where $D$ is the dilaton charge, the corresponding energy flux
density is given by
\begin{eqnarray}
J_D&=&-\left(a^2x^2+2axM+2Dax+M^2-2MD+D^2\right)^{-3}
\nonumber\\&&\cdot\frac{a^3x^5D^2\left(ax-M+D\right)^2}{\pi\left(ax+M-D\right)^{4}}\frac{da}{du}.
\end{eqnarray}
It is also related to the dilaton charge $D$ and the evolution of
$a$. Provided that $D$ is not zero, there will be a flow of
matter. When $ax\gg M-D$, we have
\begin{eqnarray}
J_D&\simeq&-\frac{D^2}{\pi a^5x^3}\frac{da}{du}.
\end{eqnarray}
We see that there is a sign difference from the flux of phantom
case Eq.(32). Thus in the expanding Universe, the dilaton black
hole accretes the surrounding matter while phantom black hole
scatters the surrounding matter. Eq.(34) can also be used to
describe a charged massive object in the FRW universe. For
radiation field $a\propto u^{{1/2}}$ or cold matter $a\propto
u^{2/3}$, the flow is towards the object all the time. This is
because the dilaton charge contributes also an attractive force.
Eq.(34) reveals that $J_D\propto Ha^{-4}$. So it follows that the
inward flow might be so great in the early universe that
super-massive black holes may be produced in a very short time.
Since we always have $D^2\geq P^2$, the overall flow of matter is
inward.
\section{black holes with both phantom and dilaton}
\hspace*{3.5mm}For completeness, we now give the solution of black
holes in the presence of both phantom and dilaton. For simplicity,
we omit the potentials of phantom and dilaton. Thus consider the
following action
\begin{eqnarray}
S&=&\int{d^4x\sqrt{-g}}\left[R-2\partial_{\mu}\phi\partial^{\mu}{\phi}+2\partial_{\mu}\psi\partial^{\mu}{\psi}
\right.\nonumber
\\ &&\left. -e^{-{2\alpha\phi}+{2\beta\psi}}F^2\right],
\end{eqnarray}
where $\phi$ and $\psi$ are for the dilaton and phantom fields,
respectively. $\alpha$ and $\beta$ are two coupling constants. We
can check that the action covers the theories of both phantom and
dilaton.\\
\hspace*{7.5mm}Varying the action with respect to the metric,
Maxwell, phantom and dilaton fields, respectively, yields
\begin{eqnarray}
0&=&\nabla_{\mu}\left(e^{-2\alpha\phi+2\beta\psi}F^{\mu\nu}\right),
\nonumber\\
 0&=&\nabla^2\phi+\frac{\alpha}{2}e^{-2\alpha\phi+2\beta\psi}F^2,
\nonumber\\
0&=&\nabla^2\psi+\frac{\beta}{2}e^{-2\alpha\phi+2\beta\psi}F^2,\nonumber\\
R_{\mu\nu}&=&2\nabla_{\mu}\phi\nabla_{\nu}\phi-2\nabla_{\mu}\psi\nabla_{\nu}\psi
\nonumber\\
&&+2e^{-2\alpha\phi+2\beta\psi}\left(F_{\mu\alpha}
F_{\nu}^{\alpha}-\frac{1}{4}g_{\mu\nu}F^2\right),
\end{eqnarray}
\hspace*{7.5mm}The most general form of the metric for the static
space-time can be written as
\begin{equation}
ds^2=-U\left(r\right)dt^2+\frac{1}{U\left(r\right)}dr^2+f\left(r\right)^2d\Omega_2^2.
\end{equation}
With the metric of Eq.(39), the equations of motion reduce to four
independent equations
\begin{eqnarray}
\frac{1}{f^2}\frac{d}{dr}\left(f^2U\frac{d\phi}{dr}\right)&=&\alpha
e^{2\alpha\phi-2\beta\psi}\frac{Q^2}{f^4},\nonumber\\
\frac{1}{f^2}\frac{d}{dr}\left(f^2U\frac{d\psi}{dr}\right)&=&\beta
e^{2\alpha\phi-2\beta\psi}\frac{Q^2}{f^4},\nonumber\\
\frac{1}{f^2}\frac{d}{dr}\left(2Uf\frac{df}{dr}\right)&=&\frac{2}{f^2}-2e^{2\alpha\phi-2\beta\psi}\frac{Q^2}{f^4},\nonumber\\
\frac{1}{f}\frac{d^2f}{dr^2}+\left(\frac{d\phi}{dr}\right)^2&=&\left(\frac{d\psi}{dr}\right)^2,
\end{eqnarray}
where $Q$ is the electric charge. The solution is obtained as
follows
\begin{eqnarray}
U&=&\left(1-\frac{r_+}{r}\right)\cdot\left(1-\frac{r_-}{r}\right)^{\frac{1-\alpha^2+\beta^2}
{1+\alpha^2-\beta^2}},\nonumber\\
f&=&r\left(1-\frac{r_-}{r}\right)^{\frac{\alpha^2-\beta^2}
{1+\alpha^2-\beta^2}},\nonumber\\
e^{2{\phi}/{\alpha}}&=&e^{2{\psi}/{\beta}}=\left(1-\frac{r_-}{r}\right)^{\frac{2}
{1+\alpha^2-\beta^2}},\nonumber\\
F^2&=&\frac{Q^2}{f^4}.
\end{eqnarray}
The two free parameters $r_+$ and $r_-$ are related to the
physical mass and charged by
\begin{eqnarray}
M&=&\frac{r_+}{2}+\frac{1-\alpha^2+\beta^2}{1+\alpha^2-\beta^2}\cdot\frac{r_{-}}{2},
\nonumber\\
Q^2&=&\frac{r_+r_{-}}{1+\alpha^2-\beta^2}.
\end{eqnarray}
When $\alpha^2=\beta^2$, the solution reduces to the
Reissner-Nordstr$\ddot{\textrm{o}}$m solution. When
$\alpha^2>\beta^2$, it is a dilaton-like black hole. On the other
hand, when $\alpha^2<\beta^2$, it is a phantom-like black hole.
There is always a sign difference before the two coupling
constants in the expressions of the metric and the physical mass
and charge. It follows once again that the effect of the phantom
field is opposite to that of the dilaton field.
\section{realization of quintom for dark energy }
Recent analysis on the properties of dark energy favor models with
the state parameter $w$ crossing $-1$ in the near past. However,
neither quintessence nor phantom can fulfill this transition. So
the models of combination of quintessence scalar field and phantom
scalar field which is called quintom are developed [23]. In this
section, we show that the quintom model can also be realized in
the dilaton-phantom frame. Consider the action in the presence of
both phantom and dilaton fields
\begin{eqnarray}
S&=&\int
d^4x\sqrt{-g}\left[R-p-2\partial_{\mu}\phi\partial^{\mu}\phi+2\partial_{\mu}\psi\partial^{\mu}\psi
\right.\nonumber
\\ &&\left. -V_1\left(\phi\right)-V_2\left(\psi\right)\right],
\end{eqnarray}
where $V_1(\phi)$ and $V_2(\psi)$ are the four dimensional
versions of equation (3) and equation (6) and $p$ is the lagrangian for the dark matter. \\
\hspace*{3.5mm}Consider a flat Universe which is described by the
flat Friedmann-Robertson-Walker metric, we can write the equations
of motion as follows
\begin{eqnarray}
3H^2&=&8\pi\left(\dot{\phi}^2-\dot{\psi}^2+\frac{1}{2}V_1+\frac{1}{2}V_2+\rho_{m0}{a^{-3}}\right),\nonumber\\
\ddot{\phi}&=&-3H\dot{\phi}-\frac{1}{4}\frac{\partial V_1
}{\partial \phi},\nonumber\\
\ddot{\psi}&=&-3H\dot{\psi}+\frac{1}{4}\frac{\partial V_2
}{\partial \psi},
\end{eqnarray}
where dot denotes the derivative with respect to $t$ and $a(t)$ is
the scale factor of the Universe. $H\equiv \dot{a}/a$ is the
Hubble parameter. $\rho_{m0}$ is the energy density of the dark
matter today. The equation of state of the dark energy is given by
\begin{eqnarray}
w=\frac{\dot{\phi}^2-\dot{\psi}^2-\frac{1}{2}V_1-\frac{1}{2}V_2}{\dot{\phi}^2-\dot{\psi}^2+\frac{1}{2}V_1+\frac{1}{2}V_2}.
\end{eqnarray}
Eq.(45) tells us if the difference of the kinetic energy between
$\phi$ field and $\psi$ field evolves, initially positive, then
zero, finally negative, then $w$ crosses $-1$ smoothly. Thus the
effect of quintom is realized. For simplicity, here we consider
the coupling constants $\alpha=1$ and $\beta=1$ and assume that
the ratio of dark matter to dark energy today is $3/7$. In Fig.1,
we plot the relation between the equation of state and the
redshift. Without the loss of generality, the initial conditions
are set $\phi(0)=0.4, \psi(0)=0.4, \dot{\phi}(0)=0,
\dot{\psi}(0)=-0.037, a(0)=1, \lambda=7$.
\begin{figure}
\begin{center}
\includegraphics[width=7.5cm]{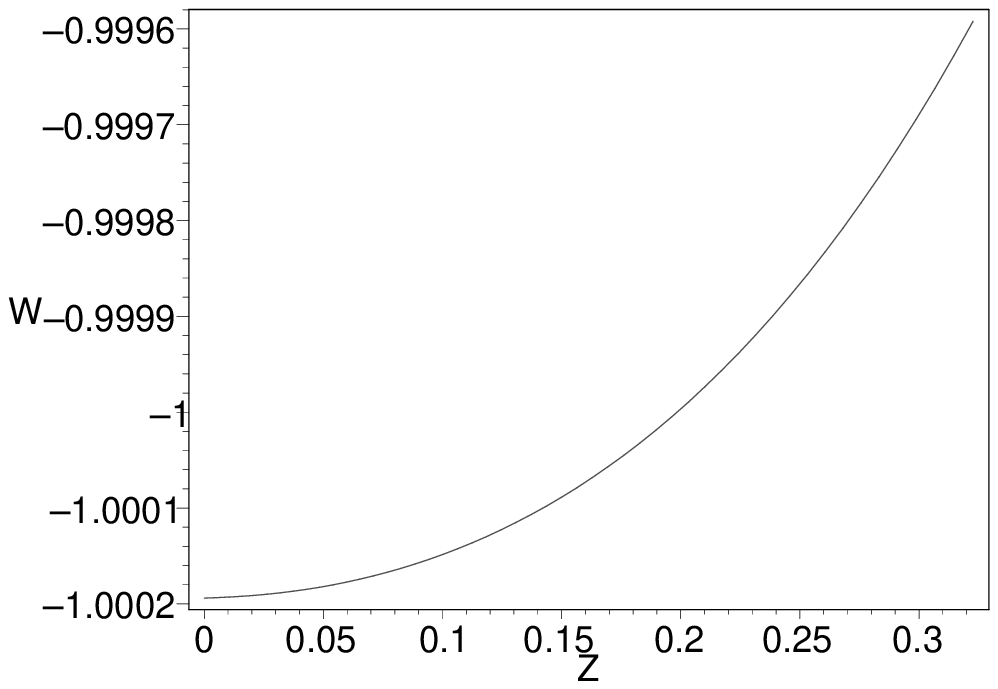}
\end{center}
\caption{$w-z$ relation.} \label{fig:wenvelope}
\end{figure}
\section{conclusion and discussion}
The dilaton action Eq.(2) corresponds to a scalar-tensor theory in
the Einstein frame, where the nonminimal coupling between scalar
and Maxwell fields arises from a conformal transformation that
brings the action from the Jordan to the Einstein frame. So the
charged dilaton black hole resembles solutions already known in
the literature. However, the phantom field considered by us is
truly different because of the negative kinetic energy. Using this
phantom field, we have constructed the exact solutions of
electrically charged phantom black holes with the cosmological
constant. The corresponding phantom potential is also obtained.
The couplings between scalars and electromagnetic fields are not
too surprising in high energy physics and they have been
considered in observations. For example, Carroll etc have studied
the astrophysical constraints on this coupling by using of the
measurements of the polarization angle and orientation of
cosmological radio sources [24]. On the other hand, Webb,
Hannestad, Anchordoqui etc have considered the astrophysical
constraints on the variation of fine structure constant using
these couplings [25].\\
\hspace*{3.5mm} We note that Bronnikov et al have investigated the
physics of neutral phantom black holes and present some
interesting results [26]. We found that the phantom field has
important consequences on the properties of black holes. For large
coupling constant, a small amount of electrical charge would make
remarkable change on the structure of spacetime. In particular,
the extremal charged phantom black holes can never
 be achieved and so the third law of thermodynamics for black holes is remedied.
Due to the phantom charge contributes an extra repulsive force to
physical mass, the phantom black hole scatters the surrounding
matter while the dilaton black hole accretes the surrounding
matter in our expanding Universe. This point is indicated once
more in the solution for black holes in the presence of both
phantom and dilaton. We also found an interior solution of a
electrically charged fluid ball immersing in the phantom field.
The solution shows that if we compress the mass $M$ and the
phantom charge $P$ in a critical radius $r_{min}$, there will need
infinite pressures at the center to against the gravity. In other
words, the ball will inevitably collapse and form a phantom black
hole. In the end, we point out that the quintom model for dark
energy can be realized in the presence of both dilaton and
phantom.
\begin{acknowledgments}
\hspace*{3.5mm}This study is supported in part by the Special
Funds for Major State Basic Research Projects, by the Directional
Research Project of the Chinese Academy of Sciences and by the
National Natural Science Foundation of China. SNZ also
acknowledges supports by NASA's Marshall Space Flight Center and
through NASA's Long Term Space Astrophysics Program.
\end{acknowledgments}

\end{document}